\documentclass[letterpaper]{article} 
\usepackage{aaai25}  
\usepackage{times}  
\usepackage{helvet}  
\usepackage{courier}  
\usepackage[hyphens]{url}  
\usepackage{graphicx} 
\usepackage{natbib}  
\usepackage{caption} 
\usepackage{algorithm}
\usepackage{algorithmic}

\usepackage{amsmath}
\usepackage{amsfonts}
\usepackage{amsthm}
\usepackage{amssymb}
\usepackage{upgreek}

\usepackage{textcomp}
\usepackage{xcolor}
\usepackage{graphicx}
\usepackage{subfig}

\usepackage{mathtools}
\usepackage{todonotes}

\newtheorem{proposition}{Proposition}
\newtheorem{theorem}{Theorem}
\newtheorem{definition}{Definition}
\newtheorem{lemma}{Lemma}
\theoremstyle{remark}
\newtheorem{remark}{Remark}

\DeclareMathOperator*{\argmin}{arg\,min}
\usepackage{newfloat}
\usepackage{listings}
\DeclareCaptionStyle{ruled}{labelfont=normalfont,labelsep=colon,strut=off} 
\floatstyle{ruled}
\newfloat{listing}{tb}{lst}{}
\floatname{listing}{Listing}
\title{Quickest Detection of Adversarial Attacks Against Correlated Equilibria}
\author{
    Kiarash Kazari,
    Aris Kanellopoulos,
    Gy\"{o}rgy D\'{a}n
}
\affiliations{
    Division of Network and Systems Engineering, School of Electrical Engineering and Computer Science\\ KTH Royal Institute of Technology, Stockholm, Sweden\\
    kkazari@kth.se, arisk@kth.se, gyuri@kth.se

}

\begin{document}

\maketitle

\begin{abstract}
We consider correlated equilibria in strategic games in an adversarial environment, where an adversary can compromise the public signal used by the players for choosing their strategies, while players aim at detecting a potential attack as soon as possible to avoid loss of utility. We model the interaction between the adversary and the players as a zero-sum game and we derive the maxmin strategies for both the defender and the attacker using the framework of quickest change detection. We define a class of adversarial strategies that achieve the optimal trade-off between attack impact and attack detectability and show that a generalized CUSUM scheme is asymptotically optimal for the detection of the attacks. Our numerical results on the Sioux-Falls benchmark traffic routing game show that the proposed detection scheme can effectively limit the utility loss by a potential adversary.    
\end{abstract}
%
 \begin{links}
     \link{Code}{https://github.com/kiarashkaz/Detection-of-Adversarial-Attacks-against-CE}
 \end{links}
\section{Introduction}
Correlated Equilibrium (CE) proposed by \cite{aumann1987_CE} is a solution concept in game theory that extends Nash Equilibria (NE) to correlated strategies among players. Unlike NE, where players choose their strategies independently, a CE represents a distribution over the space of joint actions of all players. 

Compared to NE, CE offers a more efficient and practical framework for analyzing the strategic behaviour of players. Firstly, CE accommodates scenarios that result in a higher utilitarian social welfare than what can be achieved in any NE~\cite{duffy2010_goodCE, roughgarden2002_bad_routing_equi}. Additionally, CE are computationally more tractable. As shown by \cite{chen2006_NE_complexity_1, daskalakis2009_NE_complexity_2}, computing a NE is a PPAD-complete problem even for two-player games. In contrast, a CE can be computed in polynomial time by solving a linear program~\cite{papadimitriou2008_computingCE}. Moreover, there are efficient online learning dynamics that converge to the set of correlated equilibria~\cite{cesa2006_repeated_learning_book, anagnostides2022_learning_dynamic_CE}, or to the set of coarse correlated equilibria~\cite{daskalakis2021_nearoptimalCCE, sessa2019_no_regretALg}, a related but weaker solution concept.

In a CE, coordination is facilitated by an external source of information, also called a \emph{mediator}, which helps players align their strategies. The mediator samples from a known joint distribution over the players' actions and privately communicates each player's component without revealing the others'  components. If the recommendation signals come from a CE, then no player has an interest to deviate from the mediator's suggested action, provided that the other players adhere to the recommendation. An example for a mediator is a map server in traffic routing, which recommends paths to the travelers~\cite{ning2023_routing_mediator}. Another example is recommendations issued by stock analysts for the investors in financial markets~\cite{leung2007_finance_mediator}.

A fundamental assumption for the implementation of a CE in a game is the trustworthiness of the mediator. Nonetheless, the mediator could be compromised by an adversary, with the intention to cause damage to one or more players through manipulating the signal. For instance, routing software that recommends paths to users can be vulnerable to cyber attacks. Under attack, the distribution of the recommended action profiles would deviate from the original CE. From a player's perspective, this deviation must be detected promptly, as the assumption of following recommendations is based on the distribution being known and stable. A change can impact the utilities of the players, and upon detection, players may decide to adjust their strategies accordingly. In other words, timely attack detection is crucial for ensuring situational awareness, enabling the implementation of further countermeasures after detection.  

In this paper, we demonstrate the vulnerability of the mediator's shared recommendation signal to adversarial attacks and propose a detection scheme to counteract them. We focus on the repeated play of a a CE in a strategic-form game, where an adversary manipulates the recommendation signals sent to the players by the mediator. Our key contributions are as follows:
\begin{itemize}
    \item We model the interaction between the adversary and a victim player affected by the attack as a zero-sum game;  the victim's objective is to detect the attack promptly while keeping the probability of false alarms low.
    \item We analyze the asymptotic maxmin strategies of the adversary and the defender as the cost of false alarms approaches infinity. This analysis introduces a family of distributions that enables the adversary to achieve the optimal trade-off between impact and detectability. For the defender, the optimal detection strategy is derived using the framework of quickest change detection, where the goal is to identify an optimal stopping rule.
    \item We show  the effectiveness of the proposed detection scheme in limiting the impact of the considered attacks in a  tabular game as well as in a traffic routing game on the Sioux-Falls network benchmark.
\end{itemize}

To our knowledge, the concept of adversarial attacks on correlated equilibria is unexplored in the literature. Closest to our work in the game theory literature are~\cite{roth2008_poMalice,gadjov2023_resilientNE}, which address resilience and the "price of malice" in the presence of a Byzantine player acting out of malice rather than self-interest. External attacks on a game were considered in~\cite{feng2020_attackexternal}, in the form of compromised communication channels between agents in a graph. Additionally, our work connects to \cite{lin2020_robustnessMARL, kazari2023_decentralizedMARL}, which address attack and detection in multi-agent reinforcement learning (MARL). We frame our problem as a special case of a decentralized partially-observed Markov decision problem, where state transitions are not affected by actions and where the mediator’s signal comprises players’ partial observations. Since no analytical framework exists for solving the defender-attacker interaction in the general MARL case, our approach could be seen as a first step in this direction.

The rest of the paper organized as follows: In Section \ref{sec:background}, we describe the preliminaries regarding CE and the framework of quickest change detection. Section \ref{sec:problem_formulation} presents the problem formulation. In Section \ref{sec:analyze} we analyze the best responses of the adversary and the defender and propose a detection scheme. In Section ~\ref{sec:numerical} we  evaluate the performance of the proposed detector. Section~\ref{sec:conclusion} concludes the paper.

\section{Background}
\label{sec:background}
\subsection{Correlated Equilibrium}
Consider a strategic game $\mathcal{G}=(\mathcal{N}, \lbrace \mathcal{A}^i\rbrace_{i \in \mathcal{N}}, \lbrace u^i\rbrace_{i \in \mathcal{N}})$ where $\mathcal{N}$ is the set of the players, $\mathcal{A}^i$ is player $i\in \mathcal{N}$'s action set, and $u^i(a^i,a^{-i})$ is the utility of player $i$ when she plays $a^i$ and other players play $a^{-i}$. A CE of the game is a distribution $\boldsymbol{\pi}$ over $\mathcal{A} = \bigtimes_{i \in \mathcal{N}} \mathcal{A}^i$ such that for any player $i$ and any function $\sigma^i :  \mathcal{A}^i \rightarrow \mathcal{A}^i$ we have
\begin{equation}
\label{eq:CE}
    \sum_{\mathbf{a}\in \mathcal{A}} \pi(\mathbf{a})u^i(a^i, a^{-i}) \geq \sum_{\mathbf{a} \in \mathcal{A}} \pi(\mathbf{a})u^i(\sigma^i(a^i), a^{-i}).
\end{equation}

\noindent Coarse Correlated Equilibrium (CCE) is a closely related concept, defined as a distribution $\boldsymbol{\pi}$ over $\mathcal{A} = \bigtimes_{i \in \mathcal{N}} \mathcal{A}_i$ such that for any player $i$ and any deviation $a^{i^\prime}\in\mathcal{A}^i$ we have
\begin{equation}
\label{eq:CCE}
    \sum_{\mathbf{a}\in \mathcal{A}} \pi(\mathbf{a})u^i(a^i, a^{-i}) \geq \sum_{\mathbf{a} \in \mathcal{A}} \pi(\mathbf{a})u^i(a^{i^\prime}, a^{-i}).
\end{equation}

The actual realization of a CE in a game can be obtained by introducing a mediator who picks a sample $\mathbf{s} \in \mathcal{A}$ according to $\boldsymbol{\pi}$ and recommends the corresponding $s^i$ to player $i\in \mathcal{N}$. Accordingly, player $i$'s strategy can be represented by a function $\sigma$ that maps $s^i$ to an action in $\mathcal{A}^i$. Then, (\ref{eq:CE}) can be interpreted as follows: assuming that all other players follow the mediator's recommendation, the strategy of "following the mediator's recommendation", i.e., $\sigma(s^i)=s^i$, is optimal for any player $i$. The difference between CE and CCE is that if the recommendation is drawn from a CCE, then it is a best response in expectation only before the player observes the recommendation signal. It can be shown that the set of correlated equilibria of a strategic game with finite action set is a non-empty, convex set~\cite{maschler2020_gameTheoryBook}. Moreover, the set of correlated equilibria is a subset of coarse correlated equilibria. 

\subsection{Stopping Variables and Quickest Change Detection}
Let $\mathbf{X}=\lbrace X_t, t=1,2,...\rbrace$ be a stochastic process. A stopping time $T$ with respect to $\boldsymbol{X}$ is a random variable taking values in $\lbrace 1,2,...\rbrace$ such that the event $\lbrace T=t\rbrace$ is a function of (only) $X_1,X_2,...,X_t$. 
Consider now a stochastic process $\mathbf{X}$ where observations $X_1, X_2, \ldots, X_{\nu-1}$ are independently generated from a distribution with pmf (or pdf) $f_0$, while $X_\nu, X_{\nu+1}, \ldots$ are independently generated from a different distribution $f_1$. Both $f_0$ and $f_1$ are known distributions defined on the same space, but the change point $\nu$ is unknown.
The objective of quickest change detection  is to find a stopping time $T\geq \nu$ to detect the change as quickly as possible. It is common to impose a constraint on the mean time between false alarms (MTBFA), i.e., $\mathbb{E}^{(\infty)}[T] \geq \gamma$, where $\gamma$ is a predefined threshold and $\mathbb{E}^{(\infty)}$ denotes the expectation when $\nu = \infty$, i.e., when there is no change.

There are various optimization formulations for quickest detection; one widely used formulation is to minimize the mean detection delay~\cite{lorden1971_change},
$$
W(T)=\sup_{\nu \geq 1} \text{ess sup} \mathbb{E}^{(\nu)}[(T - \nu + 1)^+ | X_1, \ldots, X_{\nu-1}],
$$
where $\mathbb{E}^{(\nu)}$ denotes the expectation when the change occurs at $\nu$, and $\text{ess sup}$ refers to the essential supremum with respect to the conditional expectation. This formulation takes into account the worst case in terms of pre-change samples.
An alternative formulation~\cite{pollak1985_optimal, lai1998_cusum_optimality_generalization} is to minimize
$\sup_{\nu \geq 1} \text{ess sup} \, \mathbb{E}^{(\nu)}[(T - \nu) \mid T \geq \nu].$

The CUSUM procedure \cite{page1954_CUSUM} is a well-known stopping time whose optimality has been shown in various cases~\cite{QCD_survey}. CUSUM is based on the Sequential Probability Ratio Test (SPRT). For a sequence of i.i.d. samples $X_1, X_2, \ldots$, the SPRT defines a stopping time to decide from which of two alternative distributions the samples $X_i$ are generated. Let  $f_1$ and $f_0$ be the pmfs (pdfs) of the two alternative hypotheses, SPRT then computes the log likelihood ratio at time $t$,
\begin{align}
    Y_t = 
        =\sum_{k=1}^t \log(\frac{f_1(X_k)}{f_0(X_k)})=\sum_{k=1}^t l(X_k),
\end{align}
and stops at the first $t$ at which $Y_t$ goes above or below pre-specified thresholds. 
CUSUM relies on the likelihood ratio test and is defined as
\begin{equation}
\label{eq:CUSUM}
T^{\text{CUSUM}}(\mu) = \inf \lbrace t: R_t= \max_{1 \leq k\leq t} \sum_{j=k}^t l(X_j) \geq \mu\rbrace,
\end{equation}
where $\mu$ is a threshold such that $\mathbb{E}^{(\infty)}[T_{\text{c}}] = \gamma$. (\ref{eq:CUSUM}) can be interpreted as applying a set of SPRT tests assuming that the change has occurred at time $k=1,2,\ldots,t$ and stop as soon as any of the likelihood ratios goes above $\mu$. Another useful representation of $R_t$ is by using the recursion
\begin{equation}
\label{eq:Cusum_recuersive}
    R_0=0, \qquad R_t=(R_{t-1}+l(X_t))^+.
\end{equation}

\section{Problem Formulation}
\label{sec:problem_formulation}
We consider a strategic game $\mathcal{G}=(\mathcal{N}, \lbrace \mathcal{A}^i\rbrace_{i \in \mathcal{N}}, \lbrace u^i\rbrace_{i \in \mathcal{N}})$, with $\mathcal{N}=\lbrace 1,2, \dots, N \rbrace$. Let $\boldsymbol{\pi}$ denote a CE of $\mathcal{G}$. We assume that $\mathcal{G}$ is played repeatedly at time steps $t=1,2,\ldots$, and at each time step, a mediator picks a sample $\mathbf{s}_t=(s^1_t, \dots, s^N_t) \sim \boldsymbol{\pi}$, and recommends $s^i_t$ to player $i\in\mathcal{N}$. Without loss of generality, we assume that $\forall i\in \mathcal{N}$ and  $\forall \mathbf{a}\in \mathcal{A}$, the utility $u^i(\mathbf{a})\geq 0$ (all  utilities in $\mathcal{G}$ can be  increased by an arbitrary constant without affecting the players' preferences). Players play a CE and all players follow the mediator's recommendation, so $a_t^i=s_t^i$. We assume that player $i$ can only observe its own utility. 

\subsection{Attack and Defense Model}
We consider a powerful adversary that manipulates the public signal $\mathbf{s}_t$ sent to the players, with the objective to reduce the utility that a particular victim player, say Player $1$, obtains in $\mathcal{G}$. 
The adversary starts to manipulate the signal at time $\nu$, and uses a stationary distribution $\boldsymbol{\tau}$ for the manipulation, i.e., the distribution $\boldsymbol{\tau}$ of adversarial manipulation at time $t\geq \nu$ is unchanged. Accordingly, $\mathbf{s}_t \sim \boldsymbol{\tau}$ for $t\geq \nu$.  At the same time, the victim player aims to detect a potential manipulation of the signal based on the sequence of utilities  $U_1, U_2, U_3, \ldots$ it observes, using an appropriately chosen stopping rule. From now on, we use the terms "Player 1" and "the defender" interchangeably. The adversary should thus remain undetectable as much as possible. 

Note that with this attack and defense model, we can assume that different action profiles result in distinct utilities for the defender. This means that for any $a^1_{(1)}$, $a^1_{(2)} \in \mathcal{A}^1$ and any $a^{-1}_{(1)}$, $a^{-1}_{(2)}\in \mathcal{A}^{-1}$, we have $u^1(a^1_{(1)},a^{-1}_{(1)})\neq u^1(a^1_{(2)},a^{-1}_{(2)})$, where we use the game theoretic notation $\mathcal{A}^{-1}$ to denote the action set of all players but Player 1. This assumption can be made because action profiles with the same utility can be grouped and treated as a single outcome. From the defender's perspective, these action profiles are indistinguishable, and from the adversary's perspective, there is no incentive to alter the distribution over them.  

\subsection{Attacker-Defender Game}

Observe that we can model the interaction between the attacker and the defender as a two-person zero-sum game $\mathcal{G}_S$. The defender chooses a stopping rule, which stops the game based on the sequence of observations. The adversary chooses a manipulation strategy $\boldsymbol{\tau}$ and a start time $\nu$. The defender's cost (or the adversary's utility) is defined based on the reduction in the total utility that Player 1 obtains in $\mathcal{G}$, as well as the cost of potential false alarms. 
In the game, Player 1's utilities $U_1, U_2, U_3, \ldots$ are the observations of the defender. Hence, for any $t \in \lbrace 1,2,\ldots\rbrace$, $U_t\in \mathcal{U}=\lbrace u^1(\mathbf{a}): \ \mathbf{a}\in \mathcal{A}\rbrace$. Note that Player 1 does not change  her strategy in $\mathcal{G}$; she follows the recommended public signal. Player 1 receives $s^1_t$, plays $a^1_t = s^1_t$, and observes $u^1_t$. Based on the sequence of her utilities and assuming that all other players follow the public signal, Player 1 tries to detect a potential attack on the public signal. The attacker-defender game $\mathcal{G}_S$ can thus be defined as follows.

\paragraph{Defender's Strategy:} Let $\mathcal{T}$ be the set of all stopping times for the sequence $U_1, U_2, \ldots$. The defender chooses a $T\in \mathcal{T}$ as its strategy. 

\paragraph{Adversary's Strategy}: The adversary chooses a (possibly stochastic) start time $\nu$ via a specification $P(\nu=t| \nu \geq t, U_1,\ldots, U_{t-1})$. A formal definition of $\mathcal{V}$, the class of all possible such start times, can be found in \cite{ritov1990_cusumGame}: Let $X_1,X_2,\ldots$ be independent random variables with $X_t \sim \text{Unif}(0,1)$. $\mathcal{V}$ is the class of all random variables $\nu$ such that $\mathbb{I}\lbrace\nu=1\rbrace$ is a measurable function of $X_1$, and for $t > 1$, $\mathbb{I}\lbrace \nu=t\rbrace$ is a measurable function of $X_t$, $\mathbb{I}\lbrace \nu < t\rbrace$ and $U_1, \ldots, U_{t-1}$.

The adversary also chooses a manipulation strategy $\boldsymbol{\tau} \in \Delta^{|\mathcal{A}|-1}$, where $\Delta^{k-1}$ denotes the $(k-1)$-dimensional probability simplex in $\mathbb{R}^k$. Since we assumed that all players follow the mediator recommendations and because Player 1's utilities in $\mathcal{G}$ for any two action profiles are different, there is a one-to-one correspondence between $\mathcal{A}$ and $\mathcal{U}$. Thus, we can consider $\boldsymbol{\pi}$ and $\boldsymbol{\tau}$ to be distributions on $\mathcal{U}$ as well.  

\paragraph{Cost (Utility) Function}: We consider that a false alarm has cost $C$ and we define the defender's cost (the adversary's utility) as 
\begin{align}
\label{eq:cost_function}
c(T,\nu,\boldsymbol{\tau}) = &\mathbb{I}\lbrace T<\nu\rbrace \left [ C - \sum_{t=1}^{ T }U_t\right] \nonumber \\ 
+ &\mathbb{I}\lbrace T\geq\nu\rbrace \left[\sum_{t=\nu}^T V_t - \sum_{t=1}^{ \nu-1 }U_t\right],  
\end{align}
where $V_t=u_{\boldsymbol{\pi}}-U_t$,  and $u_{\boldsymbol{\pi}}$ is the expected per-step utility in $\mathcal{G}$ under $\boldsymbol{\pi}$, $u_{\boldsymbol{\pi}}=\sum_{\bold{a}\in\mathcal{A}} \boldsymbol{\pi}(\bold{a})u^1(\bold{a})$. $V_t$ represents the cost imposed by the attack at time $t$. This cost is the difference between the utility that Player 1 actually obtains and the utility she would have expected to obtain had there been no attack. The negative terms in (\ref{eq:cost_function}) correspond to the utility that Player 1 obtains before the change or before stopping.   

Observe that with this formulation, it might be desirable for the adversary to select a distribution $\boldsymbol{\tau}$ that leads to an infinitesimally small per-step cost $V_t$ and an infinitely large detection time, thereby maximizing the total cost. However, in a more realistic scenario, where the time horizon is finite, such a strategy is not feasible. As such, we consider that there is a lower bound $\epsilon>0$ on the expected per-step cost that the adversary wants to cause. The set of distributions the attacker can choose $\mathbf{\tau}$ from is then 
\begin{equation}
    \mathcal{D}_\epsilon=\left\lbrace \boldsymbol{\tau}\in \Delta^{|\mathcal{A}|-1}: \mathbb{E}_{\boldsymbol{a}_t \sim \boldsymbol{\tau}} \left[\frac{1}{T-\nu+1}\sum_{t=\nu}^T V_t\right]\geq \epsilon\right\rbrace.
\end{equation}
Thus, the adversary's strategy is a pair $(\nu, \boldsymbol{\tau}) \in \mathcal{V}\times \mathcal{D_\epsilon}$. 

\section{Best Response Characterization}
\label{sec:analyze}
In this section, we analyze the asymptotic behavior of the maximin strategies of the players in $\mathcal{G}_{S}$ as the false alarm cost $C \rightarrow \infty$. In this case the defender aims to keep the probability of a false alarm low.  More precisely,  $P(T<\infty|\nu=\infty)\rightarrow 0$ as $C\rightarrow \infty$, as we show later. 

We begin by defining the components that form our proposed strategies for the players.  
\begin{definition}
\label{def:tau_theta}
    Let
    \begin{equation}
        \boldsymbol{\tau}_\theta (u) = \frac{\boldsymbol{\pi}(u)\exp(-\theta u)}{\sum_{m\in\mathcal{U}}\boldsymbol{\pi}(m)\exp(-\theta m)}, \quad u\in \mathcal{U}
    \end{equation}
    Moreover, let $\theta_{\min}$ be such that $\mathbb{E}_{U\sim \boldsymbol{\tau}_{\theta_{\min}} }[(u_{\boldsymbol{\pi}}-U)]=\epsilon$. Then, for the parameter space $\Theta=[\theta_{\min}, \infty)$, we define the family of distributions $\mathcal{F}_{\Theta}\triangleq \lbrace \boldsymbol{\tau}_{\theta}: \theta\in {\Theta}\rbrace$.
\end{definition}

\begin{definition}
\label{def:GenSPRT}
\cite{lorden1971_change}
    Let $l^\theta(U)=\log(\frac{\boldsymbol{\tau}_{\theta}(U)}{\boldsymbol{\pi}(U)})$. The generalized sequential probability ratio test of $\mathcal{F}_{\Theta}$ against $\boldsymbol{\pi}$ with threshold $\mu$ for sequence $U_1, U_2, \ldots$ is defined as 
    \begin{equation}
    \label{eq:gen_ratiotest}
        \Bar{T}(\mu)= \inf\left\lbrace t\geq 1: \sup_{\theta\in\Theta} \left(\sum_{k=1}^t l^\theta(U_k)\right) > \mu\right\rbrace
    \end{equation}
\end{definition}

\begin{definition}
    \label{def:Gen_cusum}
    \cite{lorden1971_change}
    Let $\Bar{T}(\mu)$ be as defined in Definition \ref{def:GenSPRT}. We define the stopping time $T^*(\mu)$ as
    \begin{equation}
    \label{eq:Gen_cusum}
        T^*(\mu)= \min_{k>1}\lbrace \Bar{T}_k(\mu)+k-1\rbrace,
    \end{equation} 
    where $\Bar{T}_k(\mu)$ is $\Bar{T}(\mu)$ applied to the sequence $U_k, U_{k+1}, \ldots$.
\end{definition}


Our main results are in terms of two minmax theorems based on $\boldsymbol{\tau}_{\theta_{\min}}$ as the adversary's strategy and $T^*(\mu)$ as the defender's strategy.  


Loosely speaking, the general idea we pursue is as follows: If the adversary's strategy was known to the defender, then the CUSUM procedure, whose performance is directly related to the KL-divergence of $\boldsymbol{\tau}$ and $\pi$, would be the optimal choice. The distributions in $\mathcal{F}_{\Theta}$ establish the optimal trade-off between the expected imposed cost and the mentioned KL-divergence from the adversary's point of view, motivating the definition of $\mathcal{F}_{\Theta}$. Then, with a specific set of parameterized distributions, like $\mathcal{F}_\Theta$, the stopping time defined in Definition \ref{def:Gen_cusum} is asymptotically optimal. Finally, the attacker can choose the strategy that maximizes its payoff against the mentioned stopping rule. In the sequel, the claims stated above are presented and proved formally. Section ~\ref{sec:main_theorems} includes our main results.

\subsection{The Defender's Strategy}
\label{sec:def_strategy}
 First we start by restating Lorden's asymptotic optimality results on $T^*(\mu)$. Consider a sequence $U_1, U_2, \ldots$ where $U_1,U_2, \ldots, U_{\nu-1}$ are i.i.d samples from $F_0$ and $U_{\nu},U_{\nu+1},\ldots$ are i.i.d. samples from a parameterized distribution $F_\theta$ with an unknown parameter $\theta\in\Theta$. Proposition \ref{prop:sprt_to_GenCusum} shows how to construct an optimal change detection scheme from one-sided tests, which clarifies the relation between Definition \ref{def:GenSPRT} and Definition \ref{def:Gen_cusum}, and explains the motivation behind defining $T^*$ in Definition \ref{def:Gen_cusum}. Proposition \ref{prop:exponent_family} presents sufficient conditions for one-sided tests to satisfy the requirements of Proposition \ref{prop:sprt_to_GenCusum}. Note that here the optimality criteria are in Lorden's sense, i.e., minimizing the mean detection delay $$W_\theta(T)=\sup_{\nu\geq1} \text{ess sup } \mathbb{E}^{(\nu)}_\theta[(T-\nu+1)^+|U_1,...,U_{\nu-1}],$$  where $\mathbb{E}^{(\nu)}_\theta$ denotes the expectation when the change of distribution to $F_\theta$ happens at time $\nu$. In the sequel, $\mathbb{E}_\theta$ is equivalent to $\mathbb{E}^{(0)}_\theta$.

\begin{proposition}
\label{prop:sprt_to_GenCusum}
(Theorem 1 in \cite{lorden1971_change}) Assume that there exists a class of one-sided tests $\lbrace \Bar{T}^\alpha\rbrace$ such that for all $0<\alpha<1$, 
\begin{equation}
\label{eq:flase-positive_condition}
    P(\Bar{T}^\alpha<\infty|\nu=\infty)\leq \alpha
\end{equation}
Moreover, assume that for all $\theta\in \Theta$, as $\alpha\rightarrow 0$ we have 
\begin{equation}
\label{eq:log_condition}
    \mathbb{E}_\theta[ \Bar{T}^\alpha] \sim \frac{|\log \alpha |}{D_{KL}(f_\theta||f_0)}, \ \ D_{KL}(f_\theta||f_0)<\infty
\end{equation}
For $\gamma\geq 1$, let $\alpha=\gamma^{-1}$ and define $T^\gamma=min_{k>1}\lbrace  \Bar{T}^\alpha_k+k-1\rbrace$, where $ \Bar{T}^\alpha_k$ is $ \Bar{T}^\alpha$ applied to $U_k,U_{k+1},\ldots$. Then $T^\gamma$ is a stopping variable with $\mathbb{E}^{(\infty)}[T]\geq\gamma$ that minimizes $W_\theta(T)$ for all $\theta \in \Theta$ as $\gamma \rightarrow \infty$, and we have 
\begin{equation}
\label{eq:GenCusum_performance}
    \mathbb{E}_\theta[T^\gamma] \sim  \frac{\log \gamma }{D_{KL}(f_\theta||f_0)}\quad  \text{as } \gamma \rightarrow \infty.
\end{equation}
\end{proposition}

\begin{proposition}
\label{prop:exponent_family}
    Suppose $F_\theta$'s are members of an exponential family of distributions such that 
    \begin{equation}
    \label{eq:exponential_dist}
        dF_\theta(x)= \exp (\theta x - b(\theta)) dF_0(x), \ \ \ \theta \in \Theta\cup{0} 
    \end{equation}
where $\Theta=[\theta_{min}, \infty)$ is an interval in $\mathbb{R}^+$, and $b(0)=0$.  Furthermore, let us define
\begin{equation}
\label{eq:threshold_expression}
    \mu_\alpha = \log(3 (D_{KL} (f_{\theta_{min}}||f_0)+1)^2)-\log(\alpha|\log \alpha|).
\end{equation}
 Then $\Bar{T}(\mu_\alpha)$ (defined by Definition \ref{def:GenSPRT} with $l^\theta(U)=\log(\frac{f_{\theta}(U)}{f_0(U)})$ satisfies (\ref{eq:flase-positive_condition}) and (\ref{eq:log_condition}).   
\end{proposition}

\begin{proposition}
    If the adversary's strategy $\boldsymbol{\tau}$ belongs to $\mathcal{F}_{\Theta}$, then $T^*(\mu_\alpha)$ defined in Definition \ref{def:Gen_cusum} is asymptotically optimal with respect to $W_\theta(T)$ as $\alpha\rightarrow0$.
\end{proposition}
\begin{proof}
Observe that by considering $x=-u$, $f_0\equiv\boldsymbol{\pi}$, and by defining $b(\theta)\triangleq \log (\sum_{m\in\mathcal{U}}\boldsymbol{\pi}(m)\exp(-\theta m))$, the members of $\mathcal{F}_\Theta$, defined in Definition \ref{def:tau_theta}, can be represented as in (\ref{eq:exponential_dist}). Thus, Proposition \ref{prop:exponent_family} and consequently Proposition \ref{prop:sprt_to_GenCusum} can be applied to $\Bar{T}$ and $T^*$, respectively. 
\end{proof}

As (\ref{eq:GenCusum_performance}) suggests, the asymptotic performance of $T^*$ is inversely proportional to the KL-divergence of the pre-change and post-change distributions. Next, we identify the optimal trade-off between the cost and the KL-divergence of the adversary's strategy, which leads to Definition \ref{def:tau_theta}. 

\subsection{Adversarial Strategy}
\label{sec:adv_strategy}
For a given $\delta>0$, let us consider the problem faced by the adversary,
\begin{align}
\label{eq:attack_formulation}
    & \min_{\boldsymbol{\tau}} \qquad \mathbb{E}_{\boldsymbol{\tau}}[U]  \nonumber\\ 
    &s.t. \quad \boldsymbol{\tau}\in \Delta^{|\mathcal{A}|-1}, \quad D_{KL}(\boldsymbol{\tau}||\boldsymbol{\pi}) \leq \delta.
\end{align} 
In the next proposition, we show that the solution of this problem is a distribution in $\mathcal{F}_\Theta$.
\begin{proposition}
\label{prop:optimization}
Let $\mathbf{a}_{min}\triangleq \min_{\mathbf{a}}u^1(\mathbf{a})$. If $\delta< -\log\boldsymbol{\pi}(\mathbf{a}_{min})$, then, the minimizer of (\ref{eq:attack_formulation}) is $\boldsymbol{\tau}_{\theta^*}$, where $\theta^* \in (0,\infty)$ is such that $D_{KL}(\boldsymbol{\tau}_{\theta^*}||\boldsymbol{\pi})=\delta$. Moreover, if $\delta\geq -\log\boldsymbol{\pi}(\mathbf{a}_{min})$, then (\ref{eq:attack_formulation}) attains its infimum at $\boldsymbol{\tau_\theta}$ as $\theta \rightarrow \infty$.    
\end{proposition}

\begin{remark}
\label{remark: theta}
    There is a one-to-one correspondence between $\theta$, $d(\theta)\triangleq D_{KL}(\boldsymbol{\tau}_\theta||\boldsymbol{\pi})$ and also $u_\theta\triangleq\mathbb{E}_{\boldsymbol{\tau}_\theta}[U]$. In other words, as $\theta$ varies from 0 to $\infty$, $d(\theta)$ increases monotonically from 0 to $-\log\boldsymbol{\pi}(\mathbf{a}_{min})$, and $u_\theta$ decreases monotonically from $u_{\boldsymbol{\pi}}$ to $u_\text{min}$, where $u_\text{min}= u^1(\mathbf{a}_{min})$. Consequently, if $\epsilon \in (0,u_{\boldsymbol{\pi}} - u_{\text{min}})$, then $F_{\Theta}$ as defined in Definition \ref{def:tau_theta} is well-defined, and for all $\theta\in \Theta$, $u_{\boldsymbol{\pi}} - u_\theta \geq \epsilon$.
\end{remark}

\subsection{Main Results}
\label{sec:main_theorems}
We present our main results in the next two theorems, the proofs are presented in the Appendix. For $C>1$, let $\mu(C)$ be such that
\begin{equation}
\label{eq:muC_definition}
        C=u_{\boldsymbol{\pi}}\mathbb{E}^{(\infty)}[T^*(\mu(C))]+\epsilon\mathbb{E}_{\theta_{\min}}[T^*(\mu(C))],
\end{equation}
    where $\theta_{\text{min}}$ is as defined in Definition $\ref{def:tau_theta}$. 
\begin{theorem}
\label{th:maxmin_adv}    
There is a $\nu^*(C)\in \mathcal{V}$, such that
\begin{align}
    \sup_{\boldsymbol{\tau}\in\mathcal{D}_\epsilon} \sup_{\nu\in\mathcal{V}} \inf_{T\in\mathcal{T}} &\mathbb{E}[c(T,\nu,\boldsymbol{\tau})]\ \nonumber\\
    \sim \ &\mathbb{E}[c(T^*(\mu(C)),\nu^*(C),\boldsymbol{\tau}_{\theta_{min}})]\quad  \text{as }\  C\rightarrow\infty .
\end{align}
\end{theorem}

\begin{theorem}
\label{th:maxmin_def}    
With $\mu(C)$ defined in (\ref{eq:muC_definition}), we have
\begin{align}
 \inf_{T\in\mathcal{T}} \sup_{\nu\in\mathcal{V}} \sup_{\boldsymbol{\tau}\in \mathcal{F}_\Theta}  &\mathbb{E}[c(T,\nu,\boldsymbol{\tau})]\ \nonumber\\
 \sim \ &\mathbb{E}[c(T^*(\mu(C)),\nu_1,\boldsymbol{\tau}_{\theta_{min}})]\quad  \text{ as} \  C\rightarrow\infty,
\end{align}
where $\nu_1$ is defined as $P(\nu_1=1)=1$.
\end{theorem}

Theorem \ref{th:maxmin_adv} implies that, without having any prior knowledge of the defender strategy, the highest possible cost that the adversary can guarantee is achieved by choosing $\boldsymbol{\tau}_{\theta_{\text{min}}}$ as the manipulation strategy. Additionally, if the adversary chooses this strategy the best response of the defender is $T^*(\mu(C))$.  Furthermore, Theorem \ref{th:maxmin_def} characterizes the maxmin strategy from the defender's perspective. In Theorem $\ref{th:maxmin_def}$, the supremum over $\boldsymbol{\tau}$ is taken only within $\mathcal{F}_{\theta}$, meaning that if the attack is constrained to $\mathcal{F}_{\theta}$ then choosing $T^*(\mu(C))$ leads to the lowest cost that the defender can guarantee, asymptotically. However, if the adversary is aware that $T^*(\mu(C))$ is the chosen strategy, such guarantee cannot be made, and the adversary might be able to impose a higher cost by selecting some $\boldsymbol{\tau}\in \mathcal{D}_\epsilon \setminus \mathcal{F}_{\theta}$.

\begin{remark}
\label{remark:alpha_C_relation}
    For $0<\alpha<1$, define $\mu_{\alpha}$  as in (\ref{eq:threshold_expression}) and $\gamma=\alpha^{-1}$. Then $T^*(\mu(C))$ defined by ($\ref{eq:muC_definition}$) as $C\rightarrow\infty$ is equivalent to $T^*(\mu_\alpha)$ for $\alpha\rightarrow0$ (see the proof of Theorem \ref{th:maxmin_adv} in the Appendix). Moreover, we have $P(T^*(\mu_\alpha)<\infty|\nu=\infty)\rightarrow 0$, $\mathbb{E}^{(\infty)}[T^*(\mu_\alpha)]\sim\gamma$ and $\mathbb{E}_\theta[T^*(\mu_\alpha)]\sim\frac{\log\gamma}{d(\theta)}$.
\end{remark}
Compared to solving (\ref{eq:muC_definition}), Remark \ref{remark:alpha_C_relation} provides an easier method for implementing $T^*$ based on $\alpha$ (or $\gamma$) as input value that determines the threshold $\mu$ according to (\ref{eq:threshold_expression}).  

\subsection{Implementation of the Detection Mechanism}
For a practical implementation of $T^*(\mu)$, we adapt the procedure proposed by \cite{lorden1971_change}. Note that $\Bar{T}(\mu)$ in (\ref{eq:gen_ratiotest}) can be considered as stopping  at the first time for which
\begin{equation}
\label{eq:gen_ratiotest_secondform}
     \sup_{\theta \in \Theta}\left\lbrace -\theta S_t - tb(\theta)\right\rbrace> \mu,
\end{equation}
where $S_t=U_1+\dots+U_t$. Equivalently, (\ref{eq:gen_ratiotest_secondform}) can be rewritten as
\begin{equation}
\label{eq:gen_ratiotest_thirdform}
     S_t < -\inf_{\theta \in \Theta}\left\lbrace \frac{\mu}{\theta}+ t\frac{b(\theta)}{\theta} \right\rbrace .
\end{equation}

It can be shown that the infimum in (\ref{eq:gen_ratiotest_thirdform}) is attained at $\theta_{\text{min}}$ for $t\geq \frac{\mu}{d(\theta_{\text{min}})}$, and at $\theta^{(t)}$ satisfying $d(\theta^{(t)})=\frac{\mu}{t}$ for $t< \frac{\mu}{d(\theta_{\text{min}})}$. Then for a recursive implementation of $T^*(\mu)$ in (\ref{eq:Gen_cusum}) we define a set of $M=\lfloor\frac{\mu}{d(\theta_{\text{min}})}\rfloor$ thresholds as
\begin{equation}
\label{eq:dynamic_thresholds}
    z^{(k)} =  -\frac{\mu}{\theta^{(k)}}- k\frac{b(\theta^{(k)})}{\theta^{(k)}}, \quad k=1,2,\dots, M 
\end{equation}
Moreover, we define 
$
    Q^{(k)}= U_t+\dots+U_{t-k+1} , \quad k=1,\dots, \min\lbrace M,t\rbrace
$. 
Then, $T^*(\mu)$ is equivalent to the following procedure: at each step,  perform a CUSUM test against $\theta_{\text{min}}$ with threshold $\mu$ by computing $R_t$ (see (\ref{eq:CUSUM})). Additionally, we compute $Q^{(k)}$ and we stop if $Q^{(k)}<z^{(k)}$. Note that $Q^{(k)}$ can be updated recursively as 
\begin{equation}
\label{eq:Q_update}
    Q^{(k)\prime}=\left\lbrace
    \begin{array}{cc}
       U_t + Q^{(k-1)}   &,  t-t_1\geq k\\
        0 &, t-t_1<k
    \end{array}
    \right.
\end{equation}
where $Q^{(k)\prime}$ denotes the updated value of $Q^{(k)}$ after observing $U_t$, and $t_1$ is the last time that $R_{t_1}=0$  . Whenever $R_t=1$, all previous observations can be neglected and a new cycle be started by resetting $t_1$ and setting $Q^{(k)}=0$. 
\begin{algorithm}[t]
\caption{Detection Strategy}\label{alg:Defense_alg}
\begin{algorithmic}
    \REQUIRE $\epsilon, \alpha$
    \STATE \textbf{Initialization:}
    \STATE Find ${\theta_\text{min}}$ such that $u_{\theta_\text{min}}=u_{\boldsymbol{\pi}}-\epsilon$
    \STATE Compute $\mu_{\alpha}$ as in (\ref{eq:threshold_expression})
    \STATE Find $\theta^{(k)}$ such that $d(\theta^{(k)})=\frac{\mu_\alpha}{k}, \quad k=1,\dots, M$
    \STATE Compute $z^{(k)}$ as in (\ref{eq:dynamic_thresholds}), $\quad k=1,\dots, M$
    \STATE Set $R=0,\ Q^{(k)}=0, \quad k=1,\dots, M$
    \STATE \textbf{Repeated Play Procedure:}
    \WHILE{not STOP}
    \STATE Receive $s_t$, play $a_t=s_t$, and observe $U_t$
    \STATE Compute $l_t=\log(\frac{\boldsymbol{\tau}_{\theta}(U_t)}{\boldsymbol{\pi}(U_t)})$ with $\theta=\theta_{\text{min}}$
    \STATE Update $R$ with $R\leftarrow(R+l_t)^+$
    \IF{ $R > \mu_\alpha$}
    \STATE {STOP}
    \ELSIF{$R>0$}
    \STATE Update $Q^{(k)}$ as in (\ref{eq:Q_update}), $\quad k=1,\dots, M$
        \IF{$Q^{(k)}<z^{(k)}$ for any $k$}
        \STATE STOP
        \ENDIF
    \ELSE
    \STATE Reset $R=0,\ Q^{(k)}=0, \quad k=1,\dots, M$
    \ENDIF
    \ENDWHILE

\end{algorithmic}
\end{algorithm}

\section{Numerical Results}
\label{sec:numerical}
In this section we evaluate the proposed detection scheme  against several adversarial strategies in two  games. 

\subsection{Toy Example}

\begin{table}[t]
    \centering
    \begin{tabular}{r r| c c c }
            &  \multicolumn{4}{c}{ \qquad  Player 2}\\
            &  & $A_2$ & $B_2$ & $C_2$\\
        \cline{2-5}      
        & $A_1$ & 0,0 & 6,1 & 9,3\\
Player 1& $B_1$ & 1,6 & 5,5 & 4,2\\
        & $C_1$ & 3,9 & 2,4 & 7,7\\
        \cline{2-5}
    \end{tabular}
    \caption{Payoffs in the Extended Game of Chicken.}
    \label{tab:game_1}
\end{table}

\begin{figure*}[!t]
\centering
\subfloat[Mean detection delay]{\includegraphics[width=2.8in, height=2in]{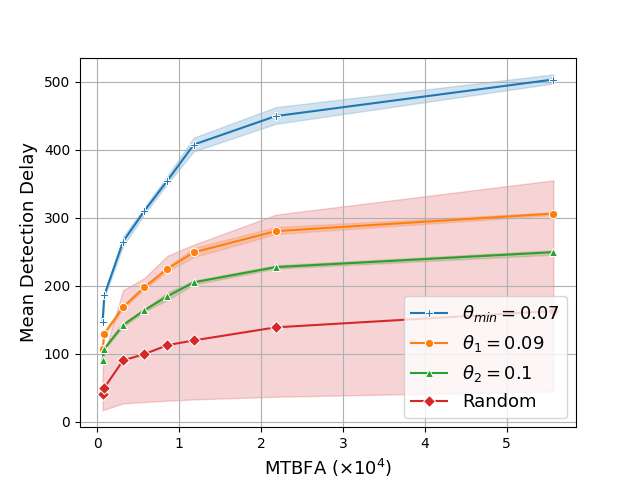}\label{fig:tabular_ttd}}
\hfil
\subfloat[Attack impact]{\includegraphics[width=2.8in, height=2in]{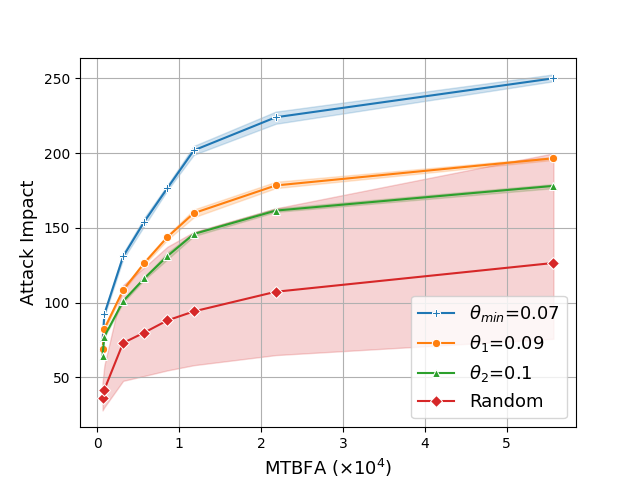}\label{fig:tabular_cost}}
\caption{Mean detection delay and attack impact with $\epsilon=0.5$ in the Extended Game of Chicken with $u_{\boldsymbol{\pi}}=6.08$. Confidence intervals are based on 5  runs.}
\label{fig:tabular}
\end{figure*}
\begin{figure*}[t]
\centering
\subfloat[Mean detection delay]{\includegraphics[width=2.8in, height=2in]{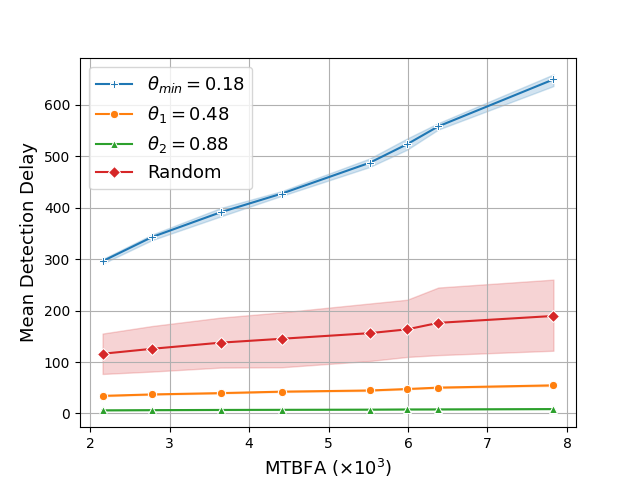}\label{fig:network_ttd}}
\hfil
\subfloat[Attack impact]{\includegraphics[width=2.8in, height=2in]{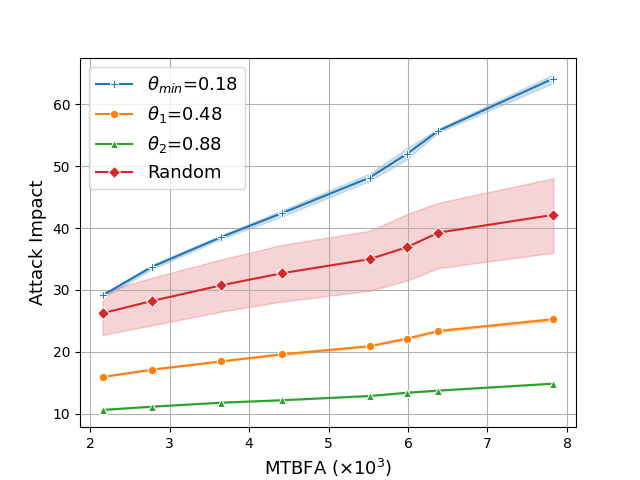}\label{fig:network_cost}}
\caption{Mean detection delay and attack impact with $\epsilon=0.1$ in the Routing Game with $u_{\boldsymbol{\pi}}=4.77$. Confidence intervals are based on 5  runs.}
\label{fig:routing}
\end{figure*}

We start with considering an extension of the Game of Chicken game~\cite{duffy2010_goodCE}, with three pure strategies.  The utilities are shown in Table~\ref{tab:game_1}. It is straightforward to verify that the following distribution is a correlated equilibrium of the game: $
   \boldsymbol{\pi}(A_1B_2)=\boldsymbol{\pi}(B_1A_2)=\boldsymbol{\pi}(B_1B_2)=\frac{1}{36}$, $\boldsymbol{\pi}(A_1C_2)=\boldsymbol{\pi}(C_1A_2)=\frac{1}{3}$, $\boldsymbol{\pi}(C_1C_2)=\frac{1}{4}$, $\boldsymbol{\pi}(A_1A_2)=\boldsymbol{\pi}(C_1B_2)=\boldsymbol{\pi}(B_1C_2)=0$. The corresponding expected pay-off for both players is $u_{\boldsymbol{\pi}}=6.08$, which results in a higher utilitarian social welfare than any pure Nash equilibrium. 
   
We consider four attack strategies. $\boldsymbol{\tau}_{\theta_{\text{min}}}$ is the maxmin strategy of the adversary with $\theta_{\text{min}}=0.07$. $\boldsymbol{\tau}_{\theta_1}$ and $\boldsymbol{\tau}_{\theta_2}$ are two other adversarial strategies with $\theta_{\text{min}}< \theta_1 = 0.09 < \theta_2=0.1$. As a baseline we use an attack where $\boldsymbol{\tau}$ is chosen randomly from the probability space $D_\epsilon$ (over elements with a non-zero probability of selection by the mediator). We refer to the different attacks as scenarios, hence we have five scenarios including the no attack scenario. As performance metrics we use the \emph{MTBFA} $\mathbb{E}^{(\infty)}[T^*]$, and the mean detection delay. Additionally, we define the \emph{attack impact} as the difference between the total utility obtained by the victim player during the period the attack is active before its detection and the expected total utility that the agent would have obtained if there had been no attack during that period. 

For the evaluation, we consider a minimum per step attack impact of $\epsilon=0.5$ and episodes with a maximum length of $10^5$ time steps.
Each episode ends when either a detection happens by the defender or the maximum number of time steps is reached. For each scenario we run 1000 episodes. In each episode of each scenario, the public signal is generated according to the corresponding adversarial strategy after the attack starts or $\boldsymbol{\pi}$ if there is no attack.  

Figure \ref{fig:tabular_ttd} shows the mean detection delay vs. MTBFA for the considered attacks, obtained by varying the detector threshold $\mu_\alpha$. The figure shows that the  time-to-detect increases as a with the logarithm of the mean time between false alerts, in accordance with Remark~\ref{remark:alpha_C_relation}. Importantly, the figure shows that the detector is robust to the attacker's choice of $\theta$; a higher value of $\theta$ allows the victim player to detect the attack sooner. Figure \ref{fig:tabular_cost} shows the attack impact during the attack as a function of the average MTBFA. The figure shows that the detection scheme effectively limits the impact the adversarial impact. For example, at an average MTBFA of $5\times10^4$ the total utility the player obtains if there is no limit on the number of time steps is around $3\times10^5$, while the attack impact with the minmax strategy is less than $250$. Furthermore, the figure indicates that $\boldsymbol{\tau}_{\theta_\text{min}}$ is the most effective attack strategy. Also, it can be observed that even with a random attack distribution that does not lie within $\mathcal{F}_{\Theta}$, the detection remains effective. 

\textbf{A note on the selection of $\epsilon$ and  $\alpha$}: In a practical scenario, the selection of $\epsilon$ and $\alpha$ is based on the player's tolerance for incurred costs and the acceptable false alarm rate. This choice is also closely tied to the time horizon over which the repeated game is intended to be played. For example, suppose that in a real scenario, the Extended Game of Chicken is to be played 500 times, and the desired false alarm rate is $\frac{1}{5\times10^4}$, which corresponds to a MTBFA of $5\times10^4$. According to Figure \ref{fig:tabular}, $\epsilon=0.5$ is a good choice for a victim player who can tolerate an adversarial attack impact of 250, i.e., around 8\% of the total expected utility $6.08\times500$ (observe that in the figure, mean detection time and the highest attack impact at an MTBFA of $5\times10^4$ are around 500, 250, respectively). In this case, if the adversary selects $\theta>\theta_\text{min}$  the total imposed cost would be lower. Also, selecting a $\theta<\theta_\text{min}$ by the adversary would result in an attack with a higher expected detection time. However, since the total length of repeated play is fixed at 500, even if the attack is not detected, the total imposed cost would still be less than 250 because the per-step cost is less than 0.5. In the Appendix, we provide additional results showing the numerical values of the attack impact as $\epsilon$ varies.

\subsection{Routing Game}

 Second, we consider a traffic routing game based on the model described in~\cite{sessa2020_contextualgames} using the Sioux Falls traffic network \cite{leblanc1975_siouxfalls_ref}. There are $N=528$ players, each aiming to transfer specific units of goods between two nodes in the network. The network comprises 24 nodes and 76 edges. Each player has $|\mathcal{A}^i|=5$ pure strategies, representing the 5 shortest paths between its source and destination. The travel time on each edge is determined by its capacity and the total volume of goods passing through it. Each player seeks to minimize the total time required to transport its goods, we thus use the negative of the total time as the utility of a player. For more details on the traffic models we refer to~\cite{sessa2020_contextualgames}. 


 A CE in this game is a distribution over $\mathcal{A}=5^{528}$ outcomes, hence infeasible to compute. We thus used  the no-regret learning algorithm GP-MW \cite{sessa2019_no_regretALg} for computing an approximate CCE $\boldsymbol{\pi}$. We selected one of the agents as the victim and quantized its utility in 1000 bins, i.e., action profiles leading to utilities within the same quantization level were considered the same. Moreover, due to the large scale of utility values, we divided all the utilities by 1000.  
 For the evaluation we ran episodes with a length of $10^4$ time steps, and used  $\epsilon=0.1$. For each scenario we ran 1000 episodes. The considered adversarial strategies correspond to $\theta_\text{min}=0.18$, $\theta_1=0.48$, $\theta_2=0.88$, and a distribution chosen randomly from probability space $D_\epsilon$. 
 
 Figure \ref{fig:routing} shows the mean detection delay (\ref{fig:network_ttd}) and the attack impact (\ref{fig:network_cost}) of the attack as a function of the MTBFA. Note that due to the limited range  of the horizontal axis (1 order of magnitude less than in Figure~\ref{fig:tabular}), the logarithmic growth of the curves is not apparent. Nonetheless, the results demonstrate that despite the complexity of the game and although the public signal is an approximate CCE, the proposed adversarial attacks on the public signal are effective. Similarly, the proposed detection method proves effective in promptly identifying the attacks and in mitigating their impact. This observation highlights the scalability of the proposed method for real-world applications. 

 \section{Conclusion}
 We considered the detection of adversarial attacks on the signals of a mediator in  correlated equilibria of non-cooperative games. We proposed a detector based on a generalized CUSUM stopping rule  designed for a specific set of adversarial distributions $\mathcal{F}_{\Theta}$. To establish the optimality of the proposed method, we modeled the interaction between the adversary and the player affected by the attack as a zero-sum game.
 We showed that any attack following a distribution in $\mathcal{F}_{\Theta}$ results in the maximum possible impact, with a constrained KL-distance to the CE, which is the primary factor determining the detection delay of the generalized CUSUM scheme. We proved that the proposed detection scheme is the maxmin strategy of the defender given that the attack distribution belongs to $\mathcal{F}_\Theta$. Additionally, we showed that the strategy that guarantees the highest possible impact for the adversary is also lies within $\mathcal{F}_\Theta$. Through numerical evaluations, we showed that the proposed method effectively mitigates the impact of various attacks, with a detection delay that increases logarithmically with the mean time between false alarms. Our results demonstrated that the detection mechanism is robust against various attacker strategies, including random attacks that fall outside $\mathcal{F}_\Theta$, and remains effective even in a complex routing game with a large number of action profiles. 
 \label{sec:conclusion}

\setcounter{secnumdepth}{0}
\section{Acknowledgments}
The work was partly funded by the Swedish Research Council through project 2020-03860 and by Digital Futures through the CLAIRE project. The computations were enabled by resources provided by the National Academic Infrastructure for Supercomputing in Sweden (NAISS) at Linköping University partially funded by the Swedish Research Council through grant agreement no. 2022-06725 .
%
%

\bibliography{ref}

\newpage
\section{Appendix}

\setcounter{secnumdepth}{2}
\setcounter{section}{0}
\renewcommand\thesection{\Alph{section}}

\section{Proof of Theorems and Propositions}
\setcounter{equation}{23}
\setcounter{proposition}{4}
\subsection{Proof of Proposition 2}

As stated in \cite{lorden1971_change}, if $\mu_\alpha$ is chosen such that $\Bar{T}(\mu_\alpha)$ satisfies (11) and $\mu_\alpha\sim |\log \alpha|$, then it satisfies (12) as well. Theorem 1 in \cite{lorden1973_open} proves that choosing $\mu_\alpha$ as mentioned in Proposition 2 satisfies (11). Moreover, we can write 
\begin{align}
    \mu_\alpha&\sim - \log(\alpha) - \log(|\log(\alpha)|)\nonumber\\
    &\sim -\log(\alpha)=|\log(\alpha)|,  \ \text{as  } \alpha \rightarrow 0
\end{align}    

\subsection{Proof of Proposition 4}

In this proof we treat $\boldsymbol{\tau}$, $\mathbf{u}$, and $\boldsymbol{\pi}$ as their representing vectors. In the following, $\bold{x}[k]$ denotes the $k$-th element of vector $\bold{x}$. (16) can be rewritten as
\begin{align}
\label{eq:attack_formulation_vector}
    &\min_{\boldsymbol{\tau}} \qquad  \bold{u}^\top \boldsymbol{\tau} \nonumber\\ 
    &s.t. \qquad ||\boldsymbol{\tau}||_1=1, \nonumber\\
    &\qquad\quad\ \ \boldsymbol{\tau} \geq 0,\quad  \nonumber \\
    & \qquad\quad D_{KL}(\boldsymbol{\boldsymbol{\tau}},\boldsymbol{\pi})= (\boldsymbol{\tau}^\top \log\boldsymbol{\tau} - \boldsymbol{\tau}^\top \log\boldsymbol{\pi})\leq \delta
\end{align} 
If $\delta=\infty$, then there is no condition on the KL-divergence and the solution is $\boldsymbol{\tau}^* = \mathbf{e}_{\mathbf{a}_{min}}$, where $\mathbf{e}$ represents the unit vector.  Otherwise we would have $D_{KL}(\boldsymbol{\tau}^*||\boldsymbol{\pi}) < \infty$, which implies that $\boldsymbol{\tau}^*[k]=0$ iff $\boldsymbol{\pi}[k]=0$. Let $\Tilde{\boldsymbol{\pi}}$ be the vector representing the non-zero elements of $\boldsymbol{\pi}$. Also, $\Tilde{\boldsymbol{\tau}}$ and $\Tilde{\bold{u}}$ represent the same indices of $\boldsymbol{\tau}$ and $\bold{u}$, respectively. Then, problem (\ref{eq:attack_formulation_vector}) can be expressed with $\boldsymbol{\pi}$, $\boldsymbol{\tau}$, and $\bold{u}$ replaced by $\Tilde{\boldsymbol{\pi}}$, $\Tilde{\boldsymbol{\tau}}$, and $\Tilde{\bold{u}}$. 

It is straightforward to verify that the optimization problem is convex and the Slater's condition holds. Accordingly, we use KKT theorem to derive optimality conditions. The Lagrangian can be written as:
\begin{align}
    \mathcal{L}=  \Tilde{\bold{u}}^\top \Tilde{\boldsymbol{\tau}}+& \eta\left((\Tilde{\boldsymbol{\tau}}^\top\log \Tilde{\boldsymbol{\tau}} - \Tilde{\boldsymbol{\tau}}^\top \log \Tilde{\boldsymbol{\pi}}) - \delta\right) \nonumber \\ 
&+  \lambda (||\Tilde{\boldsymbol{\tau}}||_1-1) - \Tilde{\boldsymbol{\beta}}^\top \Tilde{\boldsymbol{\tau}}
\end{align}
where $\eta, \Tilde{\boldsymbol{\beta}}[k] \geq 0$, and the complementary slackness implies that $\boldsymbol{\Tilde{\beta}}^*[k] \boldsymbol{\Tilde{\tau}}^*[k]=0$ and $\eta^*  (D_{KL}(\boldsymbol{\Tilde{\tau}}^*||\boldsymbol{\Tilde{\pi}}) - \delta)=0$. We know that $ \boldsymbol{\Tilde{\tau}}^*[k]\neq 0$ for any $k$, so $\boldsymbol{\Tilde{\beta}}^*[k]=0$. Setting the gradient of $\mathcal{L}$ to zero leads to 
\begin{equation}
\label{eq:lagrangian_gradient}
    \nabla_{\boldsymbol{\Tilde{\tau}}} \mathcal{L} = \Tilde{\bold{u}} + \eta (\log \boldsymbol{\Tilde{\tau}}-\log \boldsymbol{\Tilde{\pi}})  + \eta \mathbf{1} + \lambda \mathbf{1}= \bold{0}
\end{equation}
If $D_{KL}(\boldsymbol{\Tilde{\tau}}^*||\boldsymbol{\Tilde{\pi}}) < \delta$, then $\eta^*$ needs to be zero, leading to $\boldsymbol{\Tilde{{u}}} + \lambda \mathbf{1} = \bold{0}$, which is impossible. Thus, $D_{KL}(\boldsymbol{\Tilde{\tau}}^*||\boldsymbol{\Tilde{\pi}}) = \delta$. 

Solving (\ref{eq:lagrangian_gradient}) for $ \boldsymbol{\Tilde{\tau}}$, we get $\boldsymbol{\Tilde{\tau}}^*[k] = \boldsymbol{\Tilde{\pi}}[k]\exp (-\frac{\Tilde{\bold{u}}[k]}{\eta^*}-\frac{\lambda^*}{\eta^*}-1)$. The value of $\lambda^*$ can be found according to the condition $||\Tilde{\boldsymbol{\tau}}||_1=1$:
\begin{equation}
\label{eq:lambda}
       \lambda^*=\eta^*\log \left(e^{-1} \sum_k \Tilde{\boldsymbol{\pi}}[k] \exp (-\frac{\Tilde{\bold{u}}[k]}{\eta^*})\right)
\end{equation}
Finally, by substituting (\ref{eq:lambda}) into the expression of $\boldsymbol{\Tilde{\tau}}^*$, we obtain 
\begin{equation}
\label{eq: conditional_tau_expression}
    \boldsymbol{\Tilde{\tau}}^*[k] = \frac{\boldsymbol{\Tilde{\pi}}[k] \exp (-\frac{\Tilde{\bold{u}}[k]}{\eta^*})}{\sum_{k'} \boldsymbol{\Tilde{\pi}}[k'] \exp (-\frac{\Tilde{\bold{u}}[k']}{\eta^*})}
\end{equation}
By defining $\theta^* \triangleq \frac{1}{\eta^*}$, (\ref{eq: conditional_tau_expression}) takes the form of $\boldsymbol{\tau}_{\theta}$ as in Definition 1. $\theta^*$ should be such that $D_{KL}(\boldsymbol{\tau}^*,\boldsymbol{\pi})=\delta$. For the rest of the proof we require the following lemma:

\begin{lemma}
\label{lemma:d_theta_expression}
    For any $\theta$, $d(\theta)=D_{KL}(\boldsymbol{\tau}_\theta,\boldsymbol{\pi})$ can be expressed as:
    \begin{equation*}
        d(\theta)=\theta b'(\theta)-b(\theta)
    \end{equation*}
\end{lemma}
\begin{proof}
    we can write:
    \begin{align}    
    \theta b'(\theta)-b(\theta) &= \frac{-\sum_{u\in\mathcal{U}}\theta u\boldsymbol{\pi}(u)\exp(-\theta u)}{\sum_{u\in\mathcal{U}}\boldsymbol{\pi}(u)\exp(-\theta u)} - b(\theta) \nonumber \\
    &=-\sum_{u\in\mathcal{U}}\boldsymbol{\tau}_\theta(u)\theta u -\sum_{u\in\mathcal{U}}\boldsymbol{\tau}_\theta(u) b(\theta)  \nonumber \\
    & = \sum_{u\in\mathcal{U}}\boldsymbol{\tau}_\theta(u)(-\theta u-b(\theta)) \nonumber \\
    & = \sum_{u\in\mathcal{U}}\boldsymbol{\tau}_\theta(u) \log \frac{\boldsymbol{\tau}_\theta(u)}{\boldsymbol{\pi}(u)}=d(\theta)
\end{align}
\end{proof}

Note that as $\theta\rightarrow\infty$, we have $\sum_{u\in\mathcal{U}}\boldsymbol{\pi}(u)\exp(-\theta u) \sim \boldsymbol{\pi}(u_{min})\exp(-\theta u_{min})$. Accordingly, as $\theta\rightarrow\infty$:
\begin{align}
    \theta b'(\theta) = &\frac{-\sum_{u\in\mathcal{U}}\theta u\boldsymbol{\pi}(u)\exp(-\theta u)}{\sum_{u\in\mathcal{U}}\boldsymbol{\pi}(u) \exp(-\theta u)}\sim -\theta u_{min} \\
     &b(\theta)\sim \log(\boldsymbol{\pi}(u_{min})e^{-\theta u_{min}})
\end{align}
Thus, we have:
\begin{equation}
    \lim_{\theta\rightarrow\infty}d(\theta)= -\log\boldsymbol{\pi}(u_{min})
\end{equation}

Therefore, if $\delta<-\log\boldsymbol{\pi}(u_{min})$, then there exists a $\theta^*$ such that $d(\theta^*)=\delta$, and $\boldsymbol{\tau}_{\theta^*}$ is the unique solution to (16). Otherwise, the optimization problem has no solution, and the objective function reaches an infimum at $\theta=\infty$. 
\subsection{Proof of Theorem 1}
\label{sec:proof_TH1}
We can write 
\begin{align}
   \mathbb{E}[c(T,\nu,\boldsymbol{\tau})] = P( T<\nu) &\left ( C - \mathbb{E}\left[\sum_{t=1}^{ T }U_t\bigg|T<\nu\right]\right) \nonumber \\ 
   +P( T\geq\nu) &\left(\mathbb{E}\left[\sum_{t=\nu}^{ T }V_t\bigg|T \geq \nu\right]\right.  \nonumber\\
   &-\left.\mathbb{E}\left[\sum_{t=1}^{\nu-1}U_t\bigg|T\geq\nu\right]\right) 
\end{align}
Since $T$ is a stopping time, according to the Wald's identity \cite{wald1946_wald_identity}, we have
\begin{align}
    \mathbb{E}\left[\sum_{t=1}^{ T }U_t\bigg|T<\nu\right] &= \mathbb{E}[U_t\big|T<\nu]\ \cdot \mathbb{E}[T\big|T<\nu]\nonumber\\
    &=u_{\boldsymbol{\pi}}\mathbb{E}^{(\infty)}[T] \label{eq:separation_noAttack}\\
    \mathbb{E}\left[\sum_{t=\nu}^{ T }V_t\bigg|T\geq\nu\right] &= \mathbb{E}[V_t\big|T\geq\nu]\cdot\mathbb{E}^{(\nu)}_{\boldsymbol{\tau}}[T-\nu+1\big|T\geq\nu]\ \nonumber\\
    &=(u_{\boldsymbol{\pi}}-u_{\boldsymbol{\tau}})\mathbb{E}_{\boldsymbol{\tau}}^{(\nu)}[T-\nu+1\big|T\geq\nu] \label{eq:separation_attacked}
\end{align}
According to (\ref{eq:separation_noAttack}) and (\ref{eq:separation_attacked}), applying Theorem 1 in \cite{ritov1990_cusumGame} for a fixed  $\boldsymbol{\tau}$ implies that 
\begin{equation}
    \sup_{\nu\in\mathcal{V}} \inf_{T\in\mathcal{T}} \mathbb{E}[c(T,\nu,\boldsymbol{\tau})]\ = \ \mathbb{E}[c(T^{\text{CUSUM}}_{\boldsymbol{\tau}}(\mu_1(C)),\nu^*(C),\boldsymbol{\tau})]
\end{equation}
where $\mu_1(C)$ is such that 
\begin{align}
\label{eq:cusum_equality}
C &- u_{\boldsymbol{\pi}}\mathbb{E}^{(\infty)}[T^{\text{CUSUM}}_{\boldsymbol{\tau}}(\mu_1(C))] \nonumber\\
&= (u_{\boldsymbol{\pi}}-u_{\boldsymbol{\tau}}) \mathbb{E}_{\boldsymbol{\tau}}[T^{\text{CUSUM}}_{\boldsymbol{\tau}}(\mu_1(C))]\triangleq K(C,\boldsymbol{\tau}),    
\end{align}
and $\nu_*$ is defined as
\begin{equation}
\label{eq:nu_pDefinition}
    P(\nu^*=t| \nu \geq t, U_1,\ldots, U_{t-1})= p(C) (1- Q_{t-1}^{\boldsymbol{\tau}})^+.
\end{equation}
In (\ref{eq:nu_pDefinition}), $Q_t^{\boldsymbol{\tau}}=\exp(R_t^{\boldsymbol{\tau}}+l_t^{\boldsymbol{\tau}})$ (see (5)), and $p(C)$ is some probability.

$ K(C,\boldsymbol{\tau})$ is the expected post-attack cost that the adversary can guarantee. Then, choosing $\boldsymbol{\tau}$ that maximizes this cost, leads to the highest cost the adversary can guarantee. Let $\gamma_1=\mathbb{E}^{(\infty)}[T^{\text{CUSUM}}_{\boldsymbol{\tau}}(\mu_1(C))]$ .Note that as $C\rightarrow \infty$, $\mu_1(C)$ and also $\mathbb{E}^{(\infty)}(T^{\text{CUSUM}}_{\boldsymbol{\tau}}(\mu_1(C)))$ go to $\infty$. As a result, 
\begin{equation}
 \mathbb{E}_{\boldsymbol{\tau}}[T^{\text{CUSUM}}_{\boldsymbol{\tau}}(\mu_1(C))] \sim \frac{\log\gamma_1}{D_{KL}(\boldsymbol{\tau}||\boldsymbol{\pi})} , \ \ \ \text{as} \ \ C\rightarrow \infty  
\end{equation}

Thus, according to (\ref{eq:cusum_equality}), as $C\rightarrow\infty$ we have
\begin{equation}
\label{eq:C_forCUSUM}
     u_{\boldsymbol{\pi}}\gamma_1 + (u_{\boldsymbol{\pi}}-u_{\boldsymbol{\tau}}) \frac{\log\gamma_1}{D_{KL}(\boldsymbol{\tau}||\boldsymbol{\pi})} \sim C,
\end{equation} which implies that $K(C,\boldsymbol{\tau})\sim  (u_{\boldsymbol{\pi}}-u_{\boldsymbol{\tau}})\frac{\log C}{D_{KL}(\boldsymbol{\tau}||\boldsymbol{\pi})}$ as $C\rightarrow\infty$. Hence, optimal $\boldsymbol{\tau}$ for the adversary, is the one that maximizes $g(\boldsymbol{\tau})=\frac{(u_{\boldsymbol{\pi}}-u_{\boldsymbol{\tau}})}{D_{KL}(\boldsymbol{\tau}||\boldsymbol{\pi})}$.

\begin{lemma}
\label{lemma:monotonicity}
    The function $g(\theta)=\frac{u_{\boldsymbol{\pi}}-u_{\theta}}{d(\theta)}$ is monotonically decreasing in $\theta$.
\end{lemma}
\begin{proof}
We show that $g'(\theta)<0$. Note that we can write
\begin{equation}
    u_\theta= \sum_{u\in\mathcal{U}}\boldsymbol{\tau}_\theta(u)u = \frac{\sum_{u\in\mathcal{U}}u\boldsymbol{\pi}(u)\exp(-\theta u)}{\sum_{u\in\mathcal{U}}\boldsymbol{\pi}(u)\exp(-\theta u)}=-b'(\theta)
\end{equation}
Therefore, by using Lemma \ref{lemma:d_theta_expression}, we have 
\begin{align}
    g'(\theta) &= \frac{-u_{\boldsymbol{\pi}}\theta b''(\theta)}{d^2(\theta)}-\left(\frac{-b'(\theta)}{\theta b'(\theta)-b(\theta)}\right)' \nonumber\\
    &=\frac{-b''(\theta)(\theta u_{\boldsymbol{\pi}} + b(\theta))}{d^2(\theta)}.
\end{align}
$b(\theta)$ is strictly convex, and $b''(\theta)>0$ \cite{lorden1971_change}. Thus, it is sufficient to show that $\theta u_{\boldsymbol{\pi}} + b(\theta)>0$, which is equivalent to $\log(e^{\theta u_{\boldsymbol{\pi}}}\sum_{u \in \mathcal{U}}\pi(u)e^{-\theta u})>0$. Therefore, it is sufficient to show that 
\begin{equation}
\label{eq:desired}
    \sum_{u \in \mathcal{U}}\pi(u)e^{\theta ( u_{\boldsymbol{\pi}}-u)}>1
\end{equation}
Now, note that Jensen's inequality implies that
\begin{equation}
    e^{\theta \mathbb{E}_{U\sim\boldsymbol{\pi}}[u_{\boldsymbol{\pi}}-U]}\leq \mathbb{E}_{U\sim\boldsymbol{\pi}}[e^{\theta (u_{\boldsymbol{\pi}}-U)}].
\end{equation}
We have $ \mathbb{E}_{U\sim\boldsymbol{\pi}}[u_{\boldsymbol{\pi}}-U]=0$. Thus,  $\mathbb{E}_{U\sim\boldsymbol{\pi}}[e^{\theta (u_{\boldsymbol{\pi}}-U)}]\geq 1$. Since $\theta > 0$ and different values of $u\in\mathcal{U}$ are nonidentical, the equality does not hold, and we have $\mathbb{E}_{U\sim\boldsymbol{\pi}}[e^{\theta (u_{\boldsymbol{\pi}}-U)}]> 1$, which is essentially equivalent to (\ref{eq:desired}).
\end{proof}

\begin{proposition}
    \label{lemma:tau_utiliy_minmizer}
    The maximizer of $g(\boldsymbol{\tau})$ is $\boldsymbol{\tau}_{\theta_\text{min}}$.
\end{proposition}
\begin{proof}

Observe that according to Remark 1 for every $\boldsymbol{\tau}\in \mathcal{D}_\epsilon$, there exists a $\theta \in \Theta$ such that $u_{\boldsymbol{\tau}}=u_\theta$. For this $\theta$, Proposition 4 implies that $D_{KL}(\boldsymbol{\tau}||\boldsymbol{\pi})\geq D_{KL}(\boldsymbol{\tau}_\theta||\boldsymbol{\pi})$ (otherwise, $\boldsymbol{\tau}$ would have been the solution to (16)). Consequently, we have $g(\boldsymbol{\tau})\leq g(\boldsymbol{\tau}_\theta)$. As a result, the maximizer of $g(\boldsymbol{\tau})$ lies within $\mathcal{F}_\Theta$. 

Lemma \ref{lemma:monotonicity} implies that $\theta_{\text{min}}$ is the maximizer of $g(\boldsymbol{\tau}_\theta)$ within $\mathcal{F}_{\theta}$, and thus within $\mathcal{D}_{\epsilon}$.    
\end{proof}

Now suppose that the adversary chooses $(\boldsymbol{\tau}_{\theta_\text{min}}, \nu^*)$. As stated before, a Cusum stopping time with threshold $\mu_1(C)$ (defined specifically with respect to $\theta_\text{min}$) is optimal. We claim that choosing $T^*(\mu(C))$ leads to the same cost for the defender asymptotically. For $\gamma >1$,
let $\mu^*_\gamma$ be defined as in (15) with $\alpha=1/\gamma$. According to Proposition 1 as $\gamma\rightarrow \infty$, we have $\mathbb{E}^{(\infty)}[T^*(\mu^*_\gamma)]\sim \gamma$ and $\mathbb{E}_{\theta}[T^*(\mu^*_\gamma)]\sim\frac{\log\gamma}{d(\theta)}$. Consider the function $C(\gamma)\triangleq u_{\boldsymbol{\pi}}\gamma + \frac{\epsilon}{d(\theta_\text{min})}log\gamma$. $C(\gamma)$ is a one-to-one monotonically increasing function of $\gamma$. Thus, as $\gamma$ varies monotonically from 1 to $\infty$, $C(\gamma)$ varies monotonically from $u_{\boldsymbol{\pi}}$ to $\infty$. Therefore, comparing $C(\gamma)$ and (17) implies that, $T^*(\mu(C))$ as $C\rightarrow\infty$ should be the same as $T^*(\mu^*_\gamma)$ as $\gamma\rightarrow \infty$. As a result, we have
\begin{equation}
\label{eq:mu_C_solution}
  \mathbb{E}_\theta[T^*(\mu(C))]\sim \frac{\log\mathbb{E}^{(\infty)}[T^*(\mu(C))]}{d(\theta_\text{min})},  \ \ \text{as } \ C\rightarrow \infty
\end{equation}

On the other hand, applying (\ref{eq:C_forCUSUM}) for $\theta_{\text{min}}$, we get: $C\sim u_{\boldsymbol{\pi}}\gamma_1+\epsilon\frac{\log \gamma_1}{d(\theta_{\text{min}})}$ as $C\rightarrow \infty$. Hence, according to (17) and (\ref{eq:mu_C_solution}),  we must have $\mathbb{E}^{(\infty)}[T^*(\mu(C))]\sim \gamma_1$. In other words, as $C\rightarrow\infty$:
\begin{align}
      &\mathbb{E}^{(\infty)}[T^*(\mu(C))]\sim \mathbb{E}^{(\infty)}[T_\theta^{CUSUM}(\mu_1(C))] , \nonumber\\  
      &\mathbb{E}_\theta[T^*(\mu(C))]\sim \mathbb{E}_\theta[T_\theta^{CUSUM}(\mu_1(C))] 
\end{align}
Since the total cost is determined uniquely by $E^{(\infty)}[T]$ and $E_\theta[T]$, $T^*(\mu(C))$ represents a defender's best response. 

\subsection{Proof of Theorem 2}
Suppose that $\boldsymbol{\tau}_\theta\in \mathcal{F}_\Theta$ is fixed as the distribution chosen by the adversary. Let the $T_1$ be the optimal strategy for the defender, and that $\gamma=\mathbb{E}^{(\infty)}[T_1]$. Theorem 1 in \cite{lai1998_cusum_optimality_generalization} implies that
\begin{equation}
\label{eq:optimality_bound}
\sup_{\nu}\mathbb{E}_\theta^{(\nu)}[T_1-\nu+1|T_1\geq\nu]\geq\frac{\log\gamma}{d(\theta)},
\ \ \text{as} \ \ \gamma\rightarrow\infty.    
\end{equation} 

On the other hand, for $T^*(\mu_\gamma)$ defined as in Section \ref{sec:proof_TH1}, as $\gamma\rightarrow\infty$ we have $\mathbb{E}^{(\infty)}[T^*(\mu_\gamma)]\sim\gamma$, and
\begin{align}
\label{eq:optimality_achived}
  \sup_{\nu}\mathbb{E}_\theta^{(\nu)}&[T^*(\mu_\gamma)-\nu+1|T^*(\mu_\gamma)\geq\nu]\nonumber\\
  &=\mathbb{E}_\theta[T^*(\mu_\gamma)]\sim\frac{\log\gamma}{d(\theta)}  
\end{align}

As mentioned in Section \ref{sec:proof_TH1}, the total cost for the defender is a function of $\mathbb{E}^{(\infty)}[T]$ and $\mathbb{E}_\theta^{(\nu)}[T-\nu+1|T\geq\nu]$. Hence, (\ref{eq:optimality_bound}) and (\ref{eq:optimality_achived}) imply that $T^*(\mu_\gamma)$ is asymptotically the best response of the defender, as otherwise, the adversary can choose a $\nu$ that impose a higher cost. Moreover, as stated in Section \ref{sec:proof_TH1}, $T^*(\mu_\gamma)$ as $\gamma\rightarrow\infty$ is equivalent to $T^*(\mu(C))$ as $C\rightarrow\infty$. Thus $T^*(\mu(C))$ is asymptotically a best response. 

With $T^*(\mu(C))$ selected by the defender, Lemma \ref{lemma:monotonicity} directly implies that the adversary can impose the highest possible cost by choosing $\theta_\text{min}$. Moreover, $\nu=1$ corresponds to the worst-case detection delay for the defender (see (\ref{eq:optimality_achived})).  

\section{Additional Numerical Results}
\subsection{Tolerable Attack Impact}
\begin{figure}[t]
    \centering
    \includegraphics[width=\linewidth]{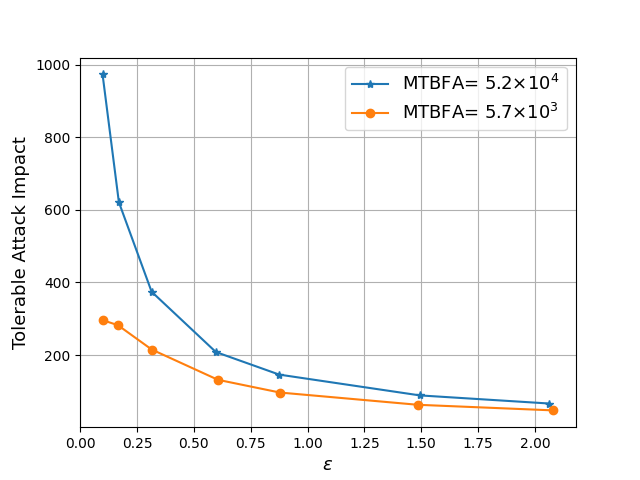}
    \caption{Tolerable attack impact as a function of $\epsilon$ for two fixed values of MTBFA in the Extended game of Chicken.}
    \label{fig:epsilon}
\end{figure}

As mentioned in Section 5.1, in a practical scenario, the selection of $\epsilon$ and $\alpha$ is based on the player's tolerance for incurred costs and the acceptable false alarm rate. This decision is influenced by the maximum cost the adversary can impose within a known time horizon. We define the tolerable cost for the defender as the impact of an attack with $\theta_{min}$, i.e., which represents the highest impact the adversary can achieve given the detector. Figure \ref{fig:epsilon} shows the tolerable attack impact for two values of MTBFA as $\epsilon$ varies. As can be observed, a higher required MTBFA (or in other words, less probability of false alarms) necessitates accepting a slightly higher attack impact. 

\end{document}